\documentclass[a4paper,12pt]{article}
\usepackage{jheppub}
\usepackage{psfrag}
\usepackage{array}
\usepackage{amssymb,amsmath,amsthm,mathtools}
\usepackage{graphicx}
\usepackage[labelsep=quad]{subcaption}
	
\usepackage{epsfig}
\usepackage{tikz}
\usepackage{hyperref}
\usepackage{color}
\usepackage{graphicx}  
\graphicspath{ {/Desktop/Dropbox/Causalityconstraints/} }
\usepackage{dcolumn}   
\usepackage{bm}        
\def\a{\alpha}
\def\b{\beta}

\def\r{\rho}
\def\s{\sigma}
\def\t{\tau}
\def\m{\mu}
\def\n{\nu}
\def\k{\kappa}
\def\th{\theta}
\def\g{\gamma}\def\G{\Gamma}
\def\L{t}\def\l{V}
\def\D{\Delta}
\def\la{\langle}
\def\ra{\rangle}
\def\o{\omega}\def\O{\Omega}
\def\d{\delta}
\def\p{\partial}

\def\oxthree{{\cal O}(x^3) }

\def\half{\textstyle{\frac{1}{2}}}
\def\bdoc{\begin{document}}
\def\edoc{\end{document}}
\def\bea{\begin{equation}}
\def\eea{\end{equation}}
\def\beq{\begin{eqnarray}}
\def\eeq{\end{eqnarray}}
\def\be{\begin{eqnarray}}
\def\ee{\end{eqnarray}}
\def\ben{\begin{enumerate}}
\def\een{\end{enumerate}}
\def\la{\langle}
\def\ra{\rangle}
\def\a{\alpha}
\def\g{\gamma}\def\G{\Gamma}
\def\d{\delta}\def\D{\Delta}
\def\e{\epsilon}
\def\z{\zeta}
\def\th{\theta}
\def\k{\kappa}
\def\l{t}
\def\m{\mu}
\def\n{\nu}
\def\o{\omega}
\def\p{\pi}
\def\r{\rho}
\def\s{\sigma}
\def\t{\tau}
\def\L{{\cal L}}
\def\S{\Sigma }
\def\gsim{\; \raisebox{-.8ex}{$\stackrel{\textstyle >}{\sim}$}\;}
\def\lsim{\; \raisebox{-.8ex}{$\stackrel{\textstyle <}{\sim}$}\;}
\def\gtrsim{\gsim}
\def\lessim{\lsim}
\def\loc{{\rm local}}
\def\vm{v_{\rm max}}
\def\bh{\bar{h}}
\def\del{\partial}
\def\nab{\nabla}
\def\half{{\textstyle{\frac{1}{2}}}}
\def\fourth{{\textstyle{\frac{1}{4}}}}
\def\bD{{\bf D}}
\def\bE{{\bf E}}
\def\bF{{\bf F}}
\def\bB{{\bf B}}
\def\bP{{\bf P}}
\def\bV{{\bf v}}
\def\bv{{\bf v}}
\def\bx{{\bf x}}
\def\by{{\bf y}}
\def\bz{{\bf z}}
\def\ba{{\bf a}}
\def\bd{{\bf d}}
\def\bs{{\bf s}}
\def\bn{{\bf n}}
\def\bp{{\bf p}}
\def\O{\Omega}
\def\br{{\bf r}}
\def\bnab{{\bf \nab}}
\def\tE{\tilde{E}}
\def\tL{\tilde{L}}
\def\Horava{Ho\v{r}ava }
\def\oxtwo{\mathscr{O}\left(x^2\right)}
\def\oxthree{\mathscr{O}\left(x^3\right)}
\def\oxfour{\mathscr{O}\left(x^4\right)}
\def\oxfive{\mathscr{O}\left(x^5\right)}
\def\LL{\text{Lanczos-Lovelock}}
\def\ph{\phantom}
\newcommand{\comm}[1]{{\bf \color{red} $\bullet$ #1 $\bullet$}}

\title{Causality constraints in Quadratic Gravity}
\author[a]{Jos\'e D. Edelstein,}
\author[b]{Rajes Ghosh,}
\author[c]{Alok Laddha}
\author[b]{and Sudipta Sarkar}
\affiliation[a]{Departamento de F\'\i sica de Part\'\i culas $\&$ Instituto Galego de F\'\i sica de Altas Enerx\'\i as (IGFAE), Universidade de Santiago de Compostela, E-15782 Santiago de Compostela, Spain.}
\affiliation[b]{Indian Institute of Technology, Gandhinagar, Gujarat 382355, India.}
\affiliation[c]{Chennai Mathematical Institute, Chennai, Tamil Nadu 603103 India.}
\emailAdd{jose.edelstein@usc.es}
\emailAdd{rajes.ghosh@iitgn.ac.in}
\emailAdd{aladdha@cmi.ac.in}
\emailAdd{sudiptas@iitgn.ac.in}
\abstract{
Classifying consistent effective field theories for the gravitational interaction has recently been the subject of intense research. Demanding the absence of causality violation in high energy graviton scattering processes has led to a hierarchy of constraints on higher derivative terms in the Lagrangian. Most of these constraints have relied on analysis that is performed in general relativistic backgrounds, as opposed to a generic solution to the equations of motion which are perturbed by higher curvature operators. Hence, these constraints are necessary but may not be sufficient to ensure that the theory is consistent. In this context, we explore the so-called CEMZ causality constraints on Quadratic Gravity in a space of shock wave solutions beyond GR. We show that the Shapiro time delay experienced by a graviton is polarization-independent and positive, regardless of the strength of the gravitational couplings. Our analysis shows that as far as the causality constraints are concerned, albeit inequivalent to General Relativity due to additional propagating modes, Quadratic Gravity is causal as per as the diagnostic proposed by CEMZ.}

\begin{document}
\maketitle
\section{Introduction}

There has been a surge of recent interest in classifying consistent classical gravitational theories in arbitrary dimensions. Until a few years ago, this question was perhaps not as appreciated in classical general relativity (GR) literature as it warrants. Apart from the constraints coming from positive energy theorems and stability of vacuum solutions, any effective field theory of gravity satisfying obvious properties such as local Lorentz invariance was considered as a viable alternative.

This notion that ``anything goes'' in classical theory was revisited in \cite{Adams:2006sv}, where the authors analysed effective field theories having meta-stable vacua. They argued that causality considerations put rather non-trivial constraints on the effective Lagrangian of non-gravitational theories. A similar constraint was also used for checking the consistency of various theories of gravity in \cite{Camanho} by Camanho, Edelstein, Maldacena, and Zhiboedov (CEMZ). They showed that the theories in which the sign of the Shapiro time shift experienced by a probe graviton depends on its polarization are acausal.

The Shapiro time shift experienced by a spinning probe in a shock-wave background, which is a classical observable \footnotetext{This statement is strictly true only in $D\, >\, 4$ dimensions. In $D\, =\, 4$ dimensions, the physical quantity is a relative time shift suffered by the probe as it passes through a pair of shock waves. For more details on this, we refer the reader to \cite{Hollowood:2016ryc} .} can be mapped to a phase shift associated with a specific scattering amplitude in the eikonal limit \cite{Camanho}. This mapping is possible because the eikonal scattering of two particles can also be viewed as the classical scattering of one particle in the background of a shock-wave sourced by the other. From this point of view, causality violation manifested in the Shapiro time advance is simply a constraint on the three-point couplings of the theory. It leads to the conclusion that any consistent theory of gravity with finitely many matter fields must not contain any non-trivial fourth and sixth derivative terms in the gravitational sector below the Planck scale. Specifically, the non-triviality makes reference to quadratic and cubic Riemann curvature terms.

However, the CEMZ construction does not put any constraint on those higher derivative terms which may have the same three-point graviton coupling as GR, but distinct four-point coupling even within this class of spacetimes. Such terms are not amenable to causality constraints probed by the sign of the Shapiro time shift, which is only sensitive to the three-point graviton couplings. Hence, a full classification of gravitational theories requires a class of consistency constraints on higher derivative terms which correspond to distinct four-point graviton interactions in perturbation theory. Since the set of all such terms is infinite-dimensional, this requirement appears to set a Herculean task. However, in a pair of interesting papers \cite{Chowdhury:2019kaq, Chandorkar:2021viw}, the authors not only classified all such higher derivative terms but also defined a set of consistency constraints dubbed as classical Regge growth (CRG). Using these constraints, they proved that in $D\, \leq\, 6$, GR is the only consistent theory. Whereas in $D\, \geq\, 7$ dimensions, there is a single term of the form $R\, \wedge\, R\, \wedge R$ satisfying the CRG criteria besides GR.

Hence, the state of the art is as follows:  A classical theory of gravity having a positive energy theorem, a stable vacuum, and massless graviton as the propagating mode in the gravitational sector is said to be `consistent' if it satisfies the causality constraints defined by CEMZ \cite{Camanho} and the CRG bound on classical $2\, \rightarrow\, 2$ scattering of gravitons \cite{Chowdhury:2019kaq}. According to these constraints, the only consistent gravitational theories with finitely many matter fields in $D\, \leq\, 6$ dimensions is General Relativity. \footnote{The case of $D=3$ is special: Theories such as New Massive Gravity and Topologically Massive Gravity were shown to be both causal and unitary \cite{Edelstein:2016nml}.} We should also remark that the CEMZ-arguments have been further scrutinized in a series of papers \cite{Papallo:2015rna, Hollowood:2015elj, deRham:2020zyh} and the idea that a time advance may imply violations of macroscopic causality in classical theories has been put into question.

Our perspective here is to accept these causality constraints on three-point graviton couplings as legitimate and explore their implications in sectors of the theory that are not perturbations around the GR regime. For the sake of concreteness, let us consider the simplest case of the so-called quadratic gravity (QG), whose Lagrangian is given in eq.(\ref{L}). In the parlance of CEMZ, this theory is not expected to be afflicted by causality issues given that it has the same three-point graviton coupling as GR. The rationale behind this argument is as follows. Solutions of QG can be mapped via field redefinition to a solution of General Relativity coupled to certain exotic matter \cite{Mozaffar:2016hmg}. Therefore, it seems that for such solutions, QG is equivalent to GR up to the field redefinition as far as three-point couplings are concerned and hence it satisfies the CEMZ constraint. However, these arguments are not enough to claim that causality is safe from transgressions in QG. We need to scrutinize the ``causal behaviour of perturbations" around an arbitrary background. The metric perturbations around the solution of QG contains massless as well as massive modes. As a result, the field redefinition of the perturbation does not yield a pure massless graviton propagating in the GR background.  Therefore, three-point graviton couplings around arbitrary solution in QG is not mapped to those of GR. Thus, the arguments presented in CEMZ for QG are necessary but they do not appear to be sufficient. It is thus an obvious question to ask whether gravitational perturbations in QG satisfies CEMZ causality constraints in a generic space-time.

In this paper, we analyse the status of causality in QG by exploring a family of shock-wave backgrounds, and compute the Shapiro time delay experienced by a probe graviton as it crosses the shock wave. We show that the sign of the Shapiro time shift experienced by a massless probe graviton is polarization independent and always positive, regardless of the value of the coupling constants. Therefore, the quadratic gravity is not afflicted by causality issues. This result consolidates the CEMZ framework  \cite{Camanho} by extending it beyond the General Relativity regime. Note that the polarization independent Shapiro time shift in shock wave backgrounds is also a feature of topologically massive gravity, and the so-called New Massive Gravity theories in $D\, =\, 3$ dimensions as well.

When viewed perturbatively around the flat (in general, maximally symmetric) spacetime, quadratic gravity is sick due to the presence of massive ghost modes. Such modes cause the Hamiltonian of linearized theory to be unbounded from below. However, a careful analysis of the full ADM Hamiltonian shows that this theory does not appear to violate the positive energy theorem \cite{Deser:2002jk}. This can be naively argued as follows. The higher derivative terms do not contribute to the ADM mass when the condition of asymptotical flatness is used. Hence, the  theory is a well-defined classical model of higher curvature gravity. Thus, although the solutions of quadratic gravity are not stable under generic perturbations, the theory does not appear to violate the positive energy theorem. In any case, due to the presence of the ghost modes, it is widely accepted that the vacuum of quadratic gravity is unstable. We insist that our interest in QG is not to consider it as a viable theory of gravity but as a toy model which may offer interesting perspective on causality constraints. So, our idea is to analyse the causality constraints for higher derivative theories of gravity which may admit non-perturbative solutions that can not be analytically continued to GR solutions. Quadratic gravity, although infected with massive ghost modes, offers the simplest test case for such a possibility.

In general, a higher-curvature theory of gravity is commonly afflicted with a myriad of inter-related problems such as causality violation,  (perturbative) instability of the vacuum, and violation of unitarity in quantum theory. Yet, it is important to classify precisely which of these properties are present in a particular higher-curvature theory of gravity since violating some of these conditions do not perhaps invalidate a theory completely; for example, CEMZ showed that unitarity is not enough to guarantee causality. Our result indicates that Quadratic Gravity is better behaved than naively expected and, thus, it may contribute towards the finer classification of consistent gravitational theories, where we need to impose restrictions on four-point (or higher) couplings in addition to the causality constraints. One such constraint is the CRG bound as depicted in \cite{Chowdhury:2019kaq, Chandorkar:2021viw}.
\section*{Review of Quadratic Gravity}
Classically any generally covariant Lagrangian is a potential starting point to define a theory of gravity. Therefore, we begin with a quadratic theory of classical gravity given by
$$
{\cal L} = \frac{1}{16 \pi G} \left( R + \alpha'\, R^2 + \beta' R_{ab} R^{ab} +\gamma' \, R_{abcd}R^{abcd} \right) ~.
$$
Note that in four dimensions, we can express the last term in terms of the other two using the Gauss-Bonnet theorem. Then, the general theory of gravity in four dimensions containing terms up to quadratic order in curvatures is given by the following Lagrangian:
\bea \label{L}
{\cal L} = \frac{1}{16 \pi G} \left( R + \alpha\, R^2 + \beta R_{ab} R^{ab} \right) ~.
\eea
This action describes what is known as \textit{Quadratic Gravity}, a theory that was extensively studied in \cite{Stelle:1976gc,Buchbinder:1992rb,Nojiri:2001ae,Alvarez-Gaume:2015rwa, Salvio:2018crh}, and whose black hole solutions are also known \cite{Berej:2006cc,Matyjasek:2004vr,Svarc:2018coe}. Interestingly, apart from the massless graviton, the theory contains a massive rank-2 tensor and a massive scalar propagating modes. These massive modes may become tachyonic unless we impose certain conditions on the higher curvature couplings. For example, in $3+1$ dimensions, the theory can be made free from the tachyonic instability if we consider the parameter space as \cite{Audretsch:1993kp}, 
\bea \label{tach}
\beta \leq 0 ~; \qquad 3 \alpha+\beta \geq 0 ~.
\eea
There are similar constraints in higher dimensions as well. Even in the parameter space where tachyonic instability is absent,  the massive spin modes are ghosts, as they have the wrong sign of the kinetic term. Such ghost fields render the vacuum of the theory unstable under linearized perturbations \cite{Salvio:2018crh} as we argued above. However, this issue will not directly concern us in the rest of the paper. In fact, the theory has been considered as an alternative theory for higher curvature gravity and it is possible to use various observational/experimental tests to constrain its dimensionful couplings \cite{Cao:2013osa,Ghosh:2019twk}.

Quadratic gravity has the same three-point graviton coupling as GR when perturbed around a flat spacetime, which is no longer true when we consider perturbations around generic backgrounds. In particular, the three-point coupling in this theory is different than GR when perturbed around a black hole solution. This implies that non-perturbative effects such as the Shapiro time delay may differ in these theories. We thus want to ask if the causality criteria of \cite{Camanho} can rule out QG, or constrain its couplings. Interestingly, as in GR \cite{'t Hooft, Sexl}, QG also admits an exact shock-wave solution which is not a solution of GR \cite{Campanelli}. Hence, to subject this theory to the CEMZ criteria, we send a probe graviton across such shock-wave. This mode will interact with the shock-wave via graviton as well as massive spin-2 ghost exchange. We shall then ask if the time delay experienced by the probe can take either sign. We will show that the resulting Shapiro delay is independent of the graviton polarization and always positive, irrespective of the values of the couplings, thereby satisfying the CEMZ causality constraint.

It may be tempting to suspect that the reason behind the positivity of the Shapiro time shift in this theory is due to the fact that solutions of quadratic are related to general relativity by a field redefinition \cite{Brigante:2007nu, Mozaffar:2016hmg}. However, we will see that this suspicion does not hold.  Although a shock-wave solution in QG theories can indeed be mapped to a shock-wave solution in General Relativity (albeit sourced by exotic matter), this field redefinition does not map a massless spin-2 perturbation on a shock wave to a massless spin-2 perturbation on a shock wave in General Relativity. We shall see, indeed, that a physical observable such as the Shapiro time delay is non-perturbative in the higher curvature coupling $\beta$. Hence there is no direct correlation between the Shapiro time delay in a generic solution of QG and  that of General Relativity.

\section{Shock wave geometry in General Relativity}

In this section we make a quick review on the propagation of a probe particle in the shock wave background of General Relativity \cite{'t Hooft, Sexl, Camanho}. The aforesaid geometry is produced by a high energy particle moving in $D$-dimensional flat spacetime. The energy-momentum tensor of such particle is given by 
\bea \label{Tu}
T_{uu}=-P_u\ \delta(u)\ \delta^{(D-2)}(\vec{x}) ~,
\eea
where $\lvert P_u \rvert$ is the associated momentum of the particle which is moving along the light-like $v$-direction ($u=0$). The coordinates $\{x_i\}$ span the transverse directions to the $uv$-plane and the quantity $r=\sqrt{x_i x^i}$ is a measure of the transverse distance. In the presence of this stress-energy tensor, we can solve the Einstein equations with the following metric-ansatz,
\bea \label{shock}
ds^2 = -du\, dv + \overline{h}_0(u, x_i)\ du^2 + \sum_i^{D-2} (dx_i)^2 ~,
\eea
and find the `profile function' $\overline{h}_0(u, x_i)$ as,
\bea \label{gr_prof}
\overline{h}_0(u, x_i) = \frac{4\ \Gamma \left(\frac{D-4}{2}\right)}{\pi^{\frac{D-4}{2}}} \ \delta(u)\ \frac{G \lvert P_u \rvert}{r^{D-4}} .
\eea
The presence of the delta-function is the evidence that the shock is localized at $u=0$. Now we consider a classical probe particle of momentum $p_v$ propagating in the shock wave background, and assume that it crosses the shock localized at $u=0$ with impact parameter $b$.
\begin{figure} [h!]
\begin{center}
\includegraphics[width=8cm]{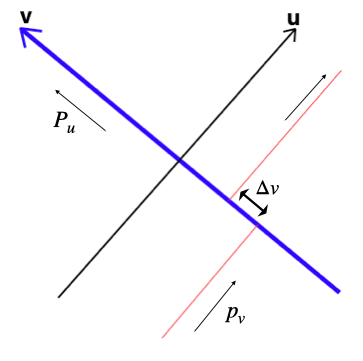}
\caption{Propagation of a probe graviton in a shock wave geometry.}
\label{figureshock}
\end{center}
\end{figure}
The delta function at $r=b$ can be lifted by introducing a new coordinate \cite{Camanho},
\bea \label{cotrans}
v = v_{\textrm{new}} +  \frac{4\ \Gamma\left(\frac{D-4}{2}\right)}{\pi^{\frac{D-4}{2}}} \ \theta(u)\ \frac{G \lvert P_u \rvert}{b^{D-4}} ~,
\eea
such that the trajectory of the probe graviton is continuous in the new coordinate. This suggests that the probe suffers a shift in the $v$-coordinate ---see Fig.\ref{figureshock},
\bea \label{gr_shift}
\Delta v_{s=0, \textrm{GR}}\,=\, \frac{4\ \Gamma\left(\frac{D-4}{2}\right)}{\pi^{\frac{D-4}{2}}}\ \frac{G \lvert P_u \rvert}{b^{D-4}} > 0 ~.
\eea
This positive shift is referred to as the Shapiro time delay.  The subscript indicates that this result is for a scalar particle scattering off the shock wave background which is a solution of Einstein's equations. Few comments on this result are in order. Note that eq.(\ref{gr_shift}) is valid only for $D>4$. For $D=4$, however, the time shift \cite{'t Hooft, Sexl} has a characteristic logarithmic dependence on $b$, {\it i.e.}, $\log(L/b)$, where $L$ is an IR cutoff \cite{CEZ}. Secondly, the deflection experienced by the probe as it crosses the shock is neglected.

One can also consider the propagation of quantum fields in the shock wave background. For this purpose, let us consider a massless scalar field $\phi$ to be the probe. In the geometry given by eq.(\ref{shock}), the wave equation takes the form
\bea \label{KG}
\del_u \del_v \phi + \overline{h}_0 \del_v^2 \phi = 0 ~.
\eea
In the above equation, derivatives in the transverse directions are neglected, since they are highly suppressed by the $u$-variation of $\overline{h}_0(u, x_i)$ from $u=0^-$ to $u=0^+$. Using  Fourier transform in $v$-direction, we can express the variation suffered by the field as follows \cite{Camanho},
\bea \label{phi_var}
\phi \left(u=0^+,\ v,\ x_i \right) = e^{-i\, p_v\, \Delta v}\, .\, \phi \left(u=0^-,\ v,\ x_i \right) ,
\eea
where the quantity $\Delta v = \int_{0^-}^{0^+} du\ \overline{h}_0(u, r=b)$, reproduces the same time delay given in eq.(\ref{gr_shift}). One can also generate the same result by performing an eikonal scattering amplitude computation in the deflectionless limit \cite{Camanho}.

\section{Graviton propagation in a shock wave in Quadratic Gravity}

Let us now discuss the shock wave solutions in higher curvature gravity theories. Note that the Lovelock theories of gravity, certainly including the Einstein-Lanczos-Gauss-Bonnet gravity, admit the same shock wave solution as in GR, (\ref{shock}) and (\ref{gr_prof}), given that their field equations remain second order, irrespective of the particular order of the higher curvature terms in the Lagrangian \cite{Horowitz, Campanelli}. The scenario changes dramatically, though, as we venture beyond the Lovelock class. Consider, for instance, a theory governed by the QG Lagrangian (\ref{L}). It is evident that the field equations obtained by varying this Lagrangian contain higher derivatives of the metric. Moreover, the domain of the unspecified parameters $\a$ and $\b$ should be constrained as in (\ref{tach}) such that there are no tachyonic modes and the correct Newtonian limit can be recovered \cite{Audretsch:1993kp}.

The theory of Quadratic Gravity admits an exact shock wave solution \cite{Campanelli}. The stress energy tensor of the source particle is the same as given in eq.(\ref{Tu}), and we again use the metric ansatz in eq.(\ref{shock}). Using the equations of motion, we obtain the `profile-function' to be,
\begin{flalign}\label{hof}
h_0(u,x_i)=f(r)\ \delta(u) ~,
\end{flalign}
where $f(r)$ is differet from that in GR and is given by \cite{Campanelli}
\begin{equation} \label{profile}
\begin{array}{ccl}
f(r) &=& -\, \displaystyle\frac{8 \pi G\, \lvert P_u \rvert\, \Gamma\left(\frac{D}{2}-1\right)}{\pi^{\frac{D}{2}-1}}\, \left[\frac{(-2 \beta)^{2 - \frac{D}{2}}}{\Gamma\left(\frac{D}{2}-1\right)} \left(\frac{r}{\sqrt{-\beta}}\right)^{2 - \frac{D}{2}}\, K_{2 -\frac{D}{2}} \left(\frac{r}{\sqrt{-\beta}}\right) \right. \\ [1em]
& & \left. \quad\qquad\qquad\qquad\qquad\qquad\qquad -\displaystyle\frac{1}{D-4}\left(\frac{1}{r}\right)^{D-4}\right] ~.
\end{array}
\end{equation}
Here, we have explicitly assumed that $\b\, \leq\, 0$ and $D > 4$. The transverse-distance is defined as before by, $r=\sqrt{x^i x_i}$, and $K_n(x)$ is the modified Bessel function of second kind. The case $D=4$ involves a logarithmic profile.
\noindent
Given the above exact shock wave background of QG, we shall now proceed to calculate the Shapiro time shift experienced by both scalar and graviton perturbations as they `cross' the shock wave. Let's first consider the case for a minimally coupled scalar particle. Via a similar calculation, it is easy to see that the Shapiro time delay experienced in the shock wave of QG is different from GR and it is given by,
\begin{flalign}
\Delta v_\textrm{s=0, \textrm{QG}}\, =\, f(b)\, .
\end{flalign}
We can also obtain the same result from the scattering amplitude computation \cite{Camanho}. As is well known, the Shapiro time shift is associated to the tree-level scattering amplitude in the eikonal limit ($t/s\, \rightarrow\, 0$) in the impact parameter representation: 
\begin{flalign}\label{july10-2}
\delta(\vec{b}, s) =\, \frac{1}{2s}\, \int \frac{d^{D-2}\, \vec{q}}{(2 \pi)^{D-2}}\, \, \, e^{i \vec{b}.\, \vec{q}}\, \, {\cal A}_{4}(s, t)\ ,
\end{flalign}
where ${\cal A}_{4}(s,t)$ is the eikonal scattering for external scalars, computed in QG. 
For massless scalar scattering and $\beta\, \leq\, 0$, the t-channel tree level scattering is given by \cite{Abe:2017abx}
\begin{flalign}
{\cal A}_{4}(s, t)\, =\, \frac{4\pi G}{t}\,\left [\, \frac{1}{-\beta\, t - 1}\, \left(\, 2 s u -\, \frac{1}{3} t^{2}\, \right)\, -\, \frac{1}{3}\, \frac{1}{-\, 2\beta t + 1}\, t^{2}\,\right ]\ ,
\end{flalign}
where $t = -\, \vert \vec{q} \vert^{2}$. In the eikonal limit the amplitude simplifies and we get,
\begin{flalign}\label{july10-1}
{\cal A}_{4}^{\textrm{eik}}(s, t)\, =\, -\, \frac{8 \pi G}{\beta}\, \frac{s^{2}}{t (t + \frac{1}{\beta})}\ .
\end{flalign}
Upon substituting the RHS of eq.(\ref{july10-1}) in eq.(\ref{july10-2}) we get (see Appendix \ref{eikonal-qg}),
\begin{flalign}\label{july12-4}
\delta(\vec{b}, s)\, =\, - p_{v}\, \triangle v_{\textrm{scalar}} \quad\Rightarrow\quad \Delta v_{\textrm{scalar}}\, =\, f(b)\, .
\end{flalign}
Next, we proceed to study the propagation of a localised gravitational perturbation in this shock wave background and calculate the time delay using the method in \cite{Camanho}. We will then provide an argument to show that the Shapiro time delay for both the scalar and the graviton is always positive.

We consider a gravitational spin-$2$ perturbation around the shock wave background $\bar{g}_{\mu \nu}$ as $g_{\mu \nu} = \bar{g}_{\mu \nu} + h_{\mu \nu}$, with $\lvert h_{\mu \nu} \rvert \ll 1$, which is localized at impact parameter $\vec{b}$ in the transverse direction. For our purpose, we only consider the traceless and transverse part of the perturbation $h_{\mu\nu}$. The equations of motion (EoM) receive contributions from the different terms in the Lagrangian \cite{Benakli:2015qlh} as listed in Table \ref{table1}, with $\Delta := - 4\left(\partial_{u} \partial_{v} + h_{0} \partial_{v}^{2}\right)$.
\renewcommand{\arraystretch}{2.5} 
\begin{table}[h!]
\begin{center}
\begin{tabular}{|c|c|}
\hline
~Terms in $\mathcal{L}$~ & ~Contribution at first order in $h_{ij}$~ \\
\hline
$R$ &	$-\frac{1}{2}\ \Delta h_{ij}$ \\
\hline
$R^2$ & $0$ \\
\hline
$R_{ab}R^{ab}$ & $-\frac{1}{2}\ \Delta^2 h_{ij}$ \\
\hline
\end{tabular}
\caption{Contribution to graviton EoM in TT-sector. }
\label{table1}
\end{center}
\end{table}
The equation of motion for the TT-sector of the graviton perturbation is therefore given by
\bea \label{EoM}
\left(\Delta+\beta \Delta^{2}\right) h_{i j} = 0 ~.
\eea
One can recover the GR limit by letting $\b \to 0^-$. Interestingly, there are no transverse derivatives in the EoM as opposed to the Einstein-Lanczos-Gauss-Bonnet case \cite{Camanho}, due to the absence of $R_{abcd} R^{abcd}$ term in the Lagrangian. Therefore, one expects that the Shapiro time delay will be independent of the choice of the graviton polarizations in QG. In the next section, we will explicitly show that is indeed the case.

\subsection{Calculation of the Shapiro time delay}
We can solve ({\ref{EoM}) exactly using the Fourier transform in $v$-direction. The most general solution can be written as (see Appendix \ref{a1}),
\bea \label{sol}
h_{i j}\left(u, v, x_{i}\right) = \int \frac{d p_v}{i p_v}\ e^{i p_v v}\ \widetilde{h}_{i j}(u, p_v, x_i) + K_{i j}^{(2)}\left(u, x_{i}\right) + C_{ij} ~.
\eea
Here tilde denotes the functions in the Fourier space of the variable $v$,
\begin{equation} \label{ht}
\begin{array}{ccl}
\widetilde{h}_{i j}\left(u, p_v, x_{i}\right) &=& \displaystyle\bigg[\ 4 i \beta p_v\ \widetilde{K}_{ij}^{(0)} \left(p_v, x_{i}\right)\ \text{exp}\left( \frac{u}{4 i \beta p_v}\right) \\ [.3em]
& & \qquad +\ \displaystyle\widetilde{K}_{i j}^{(1)} \left(p_v, x_{i}\right) \bigg]\ \text{exp}\left( -i p_v \int^{u}\!\! du\ h_{0}\left(u, x_{i}\right)\right) ~.
\end{array}
\end{equation}
Notice that there are three undetermined profile functions denoted by $\widetilde{K}_{ij}^{(m)}\left(p_v, x_{i}\right)$ and an  additive constant $C_{ij}$ in the solution. This is to be expected as we are solving a fourth order differential equation.

Since we know the full solution of the perturbation equation, one can in principle calculate the Shapiro time shift experienced by all the propagating modes in QG. However, our primary interest is  in computing  the time shift experienced  by the massless graviton and compare the result with Shapiro time shift in GR.   For this purpose, we can set $K^{(2)}_{ij}$, $C_{ij}$ to zero. Moreover, we can also drop the term associated with the profile $\widetilde{K}^{(0)}_{ij}$ for it generates the massive spin-2 mode satisfying $(1 + \beta \Delta) h_{i j} = 0$. Thus, we are left with only those solutions of the perturbation equation given by eq.(\ref{EoM}) that satisfy $\Delta\, h_{ij}= 0$. Then, the Shapiro time delay experienced by these massless probe gravitons when they `cross' the shock wave localized at $u=0$ with impact parameter $b$ is given by,
\bea \label{ts}
\Delta v = \int_{0^{-}}^{0^{+}}\!\! du\ h_{0}(u, b) = \int_{0^{-}}^{0^{+}}\!\! du\ f(b)\ \delta(u) = f(b) ~.
\eea
Note that $\Delta v$ has the same form as that of GR, the only difference being that the functional form of $f(r = b)$ has now changed from eq.(\ref{profile}), because of the presence of the higher curvature terms in the Lagrangian. This result is identical to what we have found for the scalar in eq. (\ref{july12-4}). Notice that, as per our expectation, the Shapiro time delay is independent of the graviton polarization.

We now show that the Shapiro time shift is always positive, in any arbitrary spacetime dimensions $D$, no matter the values of the quadratic couplings. To see this, let us write down eq.(\ref{ts}) in a very suggestive form,
\bea \label{Shap}
\Delta v = (\Delta v)_{\mathrm{GR}} \times \left( 1 - \frac{1}{2^{n-1} \Gamma(n)} x^{n} K_{-n}(x) \right) ~,
\eea
where, $x = b/\sqrt{-\beta}$ and $(\Delta v)_{\mathrm{GR}}$ is the Shapiro delay in General Relativity,
\bea \label{Shap_GR} 
(\Delta v)_{\mathrm{GR}} = \frac{4 \Gamma(n)}{\pi^{n}} \frac{G |P_u|}{b^{2 n}} > 0\, \, ;\, \, \qquad\mathrm{with}\ \, \, \, 2n=D-4\, .
\eea
The positivity of $\Delta v$ is now boiled down in proving the term $\left[\frac{1}{2^{n-1}\, \Gamma(n)}\, x^{n} K_{-n}(x)\right]$ is bounded from above by unity in the range $x \in (0^+, \infty)$. To prove this, let us first investigate the function, $y(x) = x^n K_{-n}(x)$ carefully.\\

\noindent
(i) Expanding about the point x=0, one gets $y(x)= 2^{n-1}\Gamma(n) + \mathcal{O}(x)$, for all real values of $n$. Also, it can be shown that, $0< y(x) < 2^{n-1}\Gamma(n)$ as $x \to 0^+$.\\

\noindent
(ii) Moreover, we have $y'(x) = \frac{d}{dx}(x^nK_{-n}(x)) = -x^nK_{-n+1}(x)$ for all real values of $n$. Thus, $y'$ is always negative if $K_{-n+1}(x)$ never changes sign in the range $x\in (0^+, \infty)$. Using property (i), $y'(0^+) < 0$, and we also know that $K_{-n+1}(x)$ has no roots in the range ($0^+,\infty $) for any real $n$. Therefore, $y'(x) < 0$ in the whole range of $x$, {\it i.e.}, $y(x) = x^n K_{-n}(x)$ is a monotonically decreasing function of $x$.\\

\noindent
(iii) $x^nK_{-n}(x) \to 0$ as $x \to \infty$.\\

\noindent
Combining the results from the aforesaid properties, one can easily check that $y(x)=x^nK_{-n}(x)$ has the maximum value of $2^{n-1}\Gamma(n)$ in the limit $x \to 0^+$, and as $x$ increases, it decreases smoothly to reach zero asymptotically in the limit $x \to \infty$. Thus, we have shown that the function $\left[ \frac{1}{2^{n-1}\Gamma(n)}\  x^n K_{-n}(x) \right]$ is bounded from above by unity in the range $x\in (0^+, \infty)$, as required.

Therefore, eq.(\ref{Shap}) suggests that $\Delta v \to 0$ as $x \to 0^+$, and it increases smoothly to match with the GR result, $\Delta v \to (\Delta v)_{\mathrm{GR}}$ in the limit $x \to \infty$ (or, $\beta \to 0^-$). This completes the proof. Rather remarkably, our result holds even for arbitrarily large negative values of the quadratic coupling $\beta$. A  few comments are in order.

As an effective field theory (EFT), Quadratic Gravity is perturbatively free of the spin-2 ghost as long as the mass of the spin-2 mode $1/\sqrt{-\beta}\, >\, \Lambda$ where $\Lambda$ is the cut-off scale that defines the domain for validity of the EFT.  We now note that in the small $\beta$ regime,  $\beta \to 0^-$ ($x \to \infty$), we have $x^nK_{-n}(x) \sim x^{n-1/2} \exp{(- x)}$. Therefore the difference between Shapiro time delay in QG and GR is exponentially suppressed. Now any detector that measures the time delay  in the EFT will have a resolution $\Delta T\, >\, \Lambda^{-1}\, \gg\, e^{-x} = e^{-b/\sqrt{-\beta}}$. Thus the time delay is indistinguishable from GR in this case.

In \cite{Camanho}, CEMZ considered the issue of fixing the causality problem in higher derivative theories (more specifically theories which are deformations of GR by Lanczos-Gauss-Bonnet term and a $R^{3}$ term),  by allowing propagation of massive spin-2 particles in the t-channel. In other words, they considered three point couplings involving one massive spin-2 particle and two massless gravitons which in the $m\, \rightarrow\, 0$ limit of the massive particle reduced to Gauss-Bonnet or $R^{3}$ couplings. It was shown that such an addition of massive propagating spin-2 mode can not fix the causality problem infecting these higher derivative couplings.

We note that our result is not in contradiction with theirs. As we saw in eq.(\ref{Shap}), the propagating massive spin-2 ghost mode reduces the Shapiro time delay as compared to the GR result. However, rather surprisingly the correction term $\frac{1}{2^{n-1} \Gamma(n)} x^{n} K_{-n}(x) (\Delta v)_{\mathrm{GR}}$ is bounded from above by $(\Delta v)_{\mathrm{GR}}$ and hence we never obtain a time advance. The underlying reason for this upper bound however is not clear to us. In fact, for a fixed mass $\frac{1}{\sqrt{-\beta}}$ of the ghost, as we decrease the impact parameter, the correction term increases and tends to $(\Delta v)_{\textrm{GR}}$ in $b\, \rightarrow\, 0$ limit. This feature (the correction from GR result being enhanced in the small impact parameter limit) is in line with what was already noticed in \cite{Camanho}.  

\subsection{Subtleties with Field Redefinition}

In this subsection, we analyse the absence of Shapiro time advance in QG in light of the fact that any exact solution of QG can be mapped onto an exact solution of GR under a field redefinition \cite{Mozaffar:2016hmg}. That is, we would like to understand if the Shapiro time delay computed in the shock-wave background in QG is equivalent to the Shapiro time delay computed in the corresponding spacetime in GR. For this purpose, we first apply the field redefinition proposed in \cite{Mozaffar:2016hmg} to the shock-wave solution of quadratic gravity. Given such a solution $g^{\mu \nu}$ to Quadratic Gravity (coupled to matter which we assume satisfies the null energy condition), the mapping to the corresponding shock-wave solution $\overline{g}^{\mu \nu}$ in General Relativity is given by the following formula:
\begin{flalign} \label{map}
\overline{g}^{\mu \nu}\, = \, g^{\mu \nu} + 2\, \beta\, R^{\mu \nu} \ ,
\end{flalign}
where we have used the fact that the shock-wave metric has vanishing Ricci scalar. Using the QG shock-wave profile given in eqs.(\ref{hof},\ref{profile}), it can be checked easily that the mapped solution $\overline{g}_{\mu \nu}$ is indeed the GR shock-wave solution of the form eq.(\ref{shock}) with the profile given by eq.(\ref{gr_prof}). However, note that as QG has more propagating modes than GR, the solution $\overline{g}_{\mu\nu}$ is sourced by the stress tensor which includes exotic contributions associated to the massive spin-2  and a massive scalar modes \cite{Mozaffar:2016hmg} in addition to the matter.

Although the shock-wave metric of QG is mapped onto the shock-wave metric in GR, the Shapiro time shift is an observable which depends on the physics of perturbations around the shock-wave background. We now argue that  massless spin-2 perturbations around the shock-wave solution in QG are not mapped onto graviton perturbations around the corresponding shock-wave in GR. It in turn confirms our assertion that the Shapiro time shift computed in shock-wave background of QG is not equal to that of GR modulo a field redefinition.

The basic idea is very simple. The field redefinition which maps $g_{\mu \nu}\, \rightarrow\, \overline{g}_{\mu \nu}$, will also map a perturbation $h_{\mu\nu}$ over the QG-shockwave to the perturbation $\overline{h}_{\mu \nu}$ over the GR-shockwave. Using a similar mapping as given in eq.(\ref{map}), it can be shown that the perturbation $\overline{h}_{\mu \nu}$ satisfies the following equation: 
\bea \label{hbar}
\overline{\Delta}\ \overline{h}_{ij}\, =\, 4\, \left[h_0 - \overline{h}_0\right]\, \partial_{v}^{2}\, (1\, +\, \beta\, \Delta)\, h_{ij}\, ~,
\eea
where $\overline{\Delta}$ and $\overline{h}_0$ are the Laplacian operator and the shock-wave profile in GR. Note also that in the shock-wave background of GR, massless spin-2 perturbations obey the equation: $\overline{\Delta}\ \overline{h}_{ij}=0$, which is in general different from the mapped equation. One way to map eq.(\ref{hbar}) to that of GR is to consider only those perturbations over the QG shock-wave which satisfy $(1\, +\, \beta\, \Delta)\, h_{ij}\, =\, 0$.

Using eq.(\ref{EoM}) one can infer that $h_{ij}$ which is in the kernel of $\Delta$-operator is the massless spin-2 perturbation. On the other hand, $h_{ij}$ which is in the kernel of $(1\, +\, \beta\, \Delta)$-operator is the massive spin-2 perturbation. Thus, from eq.(\ref{hbar}), one can say that the field re-definition maps the QG perturbation equation given by eq.(\ref{EoM}) to that of GR, {\it i.e.}, $\overline{\Delta}\, \overline{h}_{ij}\, =\, 0$, iff $h_{ij}$ is the massive spin-2 mode. However, for the massless spin-2 perturbation, the Shapiro time delay is not equal to the corresponding quantity in General Relativity. 

\section{Generalization to higher order theories}

The structure of eq.(\ref{EoM}) encourages us to generalize the above result for a special class of theories classified by a graviton EoM of the following form (with $n \in Z^+$),
\bea \label{gen_eom}
\left( 1 + \gamma \Delta^n \right) (\Delta h_{ij}) = 0 ~,
\eea
where $\Delta = -4 (\partial_u\partial_v + h_0 \partial_v^2)$. This structure is special because of two distinct features:\\

\noindent
(i) The differential operator acting on the perturbation $h_{ij}$ supports factorization into two parts: the GR part ($\Delta$) and the remaining factor, $(1 + \gamma \Delta^n)$, coming from the higher curvature terms (with coupling constant $\gamma$) of the Lagrangian.\\

\noindent
(ii) The EoM does not contain any transverse derivatives of the profile function $h_0$.\\

One can show that the theories which give rise to such an EoM also produce the Shapiro time shift of an amount $\Delta v= f(b)$. However, the form of $f(r)$ will vary for different theories. Thus, depending upon the sign of $f(b)$, we can fix the sign of the time shift, too. One such theory is given by the Lagrangian
\begin{equation}\label{Lag}
\mathcal{L} = \sqrt{-g} (R + \alpha R^2 + \gamma R_{ab} \nabla_c \nabla^c R^{ab}) ~.   
\end{equation}
It leads to the graviton EoM of the form eq.(\ref{gen_eom}) with $n=2$, provided that it supports a shock wave solution to begin with.
\section{Discussions and Conclusion}
One of the remarkable results in \cite{Camanho}  was to identify the Shapiro time shift in a shock wave background with the eikonal Phase in a $2\, \rightarrow\, 2$ scattering amplitude in a generic theory of gravity.\footnote{This result was of course well known for General Relativity since the work of \cite{'t Hooft, Kabat}.} Naively one may think that as three point graviton coupling (in flat space background) is the same in QG as in GR, the tree-level eikonal scattering of scalars or gravitons at fixed impact parameter will be equal in the two theories. However, as QG has additional propagating modes, namely a massive spin 2 ghost, the tree-level $2\, \rightarrow\, 2$ scattering amplitude is different in the two theories.  As we verified explicitly in the case of Shapiro time shift of the scalar particles, it is the propagating ghosts which non-trivially modify the time delay.   Hence we expect that the phase shift computation using tree-level 4 graviton scattering will also reproduce the classical computation for the time shift experienced by a probe graviton. However as four graviton scattering in quadratic gravity is a rather complicated beast, we have not verified this assertion explicitly.

The minimal list of necessary conditions that a consistent gravitational effective field theory must obey include (i) the existence of a stable vacuum, (ii) the validity of a positive energy theorem, and (iii) the causality properties as advocated and quantified by CEMZ in \cite{Camanho}. From this perspective, Quadratic Gravity, although not equivalent to General Relativity via mere field redefinitions, seems to be better behaved than naively expected. It seems to be a representative of a class of theories which deserve further attention. One way to rule (this class of) theories out is by demanding perturbative stability of the vacuum, and this clearly renders QG inconsistent. However this is not  an obvious criterion for isolating inconsistent gravitational theories. In non-gravitational theories perturbative instability is tied to instability of the vacuum, which is no longer obvious in a gravitational theory. So although perturbation theory is ill-defined due to presence of massive ghosts, the positive energy theorem is not violated due to the existence of higher curvature terms and hence the analysis of stability of vacuum needs further scrutiny. Another possiblity is to consider QG as an effective field theory (with no sensible UV completion) but perturbative unitarity upto a cut-off scale which is below the mass of the ghost. But in this case, QG is \emph{indistinguishable} from GR as far as Shapiro time delay is concerned. 

Nevertheless, we know that there may be further necessary conditions that one has to impose to classify consistent gravitational and gauge theories. As mentioned previously, one such condition is the CRG conjecture formulated in \cite{Chowdhury:2019kaq}. CRG conjecture analyses consistency of classical theories in terms of the tree-level scattering amplitudes and states that in the Regge limit, the classical scattering amplitudes are bounded by $s^{2}$ in the large $s$ limit. General Relativity saturates this bound.

There are many motivations for the CRG conjecture including bounds on the growth of the S-matrix in the t-channel which are satisfied by all the known consistent theories \cite{Camanho}. It has been proved rigorously in AdS background using  certain ideas in holography,  \cite{Chandorkar:2021viw}. There has also been a proof in flat background by analysing scattering amplitude in flat space as a limit of AdS amplitudes \cite{Caron-Huot:2021enk}. Hence, the conjecture is on a firm footing and  adds to the list of the set of consistency conditions for gravitational theories in any dimension. It will be interesting to analyse the four graviton scattering in QG and verify whether it satisfies the CRG bound. We leave this question for the future.

\section*{Acknowledgements}

We would like to thank  Subham Dutta Chowdhary, Juan Maldacena, Shiraz Minwalla, Bayram Tekin, Miguel \'Angel V\'azquez Mozo and Alexander Zhiboedov for insightful discussions and their comments.
The work of JDE is supported by MINECO FPA2017-84436-P, Xunta de Galicia ED431C 2017/07, Xunta de Galicia (Centro singular de investigaci\'on de Galicia accreditation 2019-2022), the European Union (European Regional Development Fund -- ERDF), the ``Mar\'\i a de Maeztu'' Units of Excellence MDM-2016-0692, and the Spanish Research State Agency.
The research of RG is supported by the Prime Minister Research Fellowship (PMRF-192002-120), Government of India.  AL would like to thank IIT Gandhinagar scholar in residence program for their hospitality during the initial stage of this project.  The research of SS is supported by the Department of Science and Technology, Government of India under the SERB CRG Grant (CRG/2020/004562). 

\appendix

\section{Shapiro Time Shift from eikonal scattering}
\label{eikonal-qg}

In this appendix, we derive the Shapiro time delay suffered by a probe scalar using tree-level scattering amplitude in the eikonal limit. 
The relationship between the Shapiro time shift and the phase-shift obtained from tree-level amplitude in eikonal limit is given by,
\begin{flalign}\label{july12-3}
\Delta v\, =\, -\, \frac{1}{p_{v}}\, \delta(\vec{b}, s)\ .
\end{flalign}
The phase shift $\delta(\vec{b}, s)$ is proportional to the tree-level amplitude in impact parameter representation, see eq.(\ref{july10-2}). Using the $2\, \rightarrow\, 2$ scalar amplitude given by eq.(\ref{july10-1}), we can use the technique of partial fraction decomposition to rewrite the phase shift as follows,
\begin{flalign}
\delta(\vec{b}, s)\, =\, -\, 4\pi G s\, \int \frac{d^{D-2} \vec{q}}{(2 \pi)^{D-2}}\, \, e^{i \vec{b} \cdot \vec{q}}\, \left[\frac{1}{t}\, -\, \frac{1}{(t + \frac{1}{\beta})} \right]\ .
\end{flalign}
Using $t=-\lvert \vec{q} \rvert^2$, and  $b\, =\, (\vec{b} \cdot \vec{b})^{\frac{1}{2}}$, one can produce the following results,
\begin{flalign}
\int \frac{d^{D-2} \vec{q}}{(2 \pi)^{D-2}}\, e^{i \vec{b} \cdot \vec{q}}\, \, \frac{1}{t}\, =\, -\, \frac{1}{4 \pi^{D/2-1}}\, \, \frac{\Gamma[D/2-2]}{b^{D-4}}
\end{flalign}
Similarlly,\\
\begin{flalign}
\int \frac{d^{D-2} \vec{q}}{(2 \pi)^{D-2}}\, e^{i \vec{b} \cdot \vec{q}}\, \, \frac{1}{t + \frac{1}{\beta}}\, =\, -\, \frac{(b \sqrt{-\beta})^{2-D/2}}{\left( 2 \pi \right)^{D/2-1}}\, \, K_{2-D/2} \left[\frac{b}{\sqrt{-\beta}}\right]\ .
\end{flalign}
While evaluating the integrals, we have explicitly used the condition $\beta\, \leq\, 0$. Combining these results and substituting $s\, =\, -4\, \vert P_{u}\vert\, p_{v}$, one can verify eq.(\ref{july12-4}). 

\section{Perturbation in Quadratic Gravity}\label{a1}
\label{App}

Now, we outline the method for deriving the solution of the graviton equation of motion (\ref{EoM}). For this purpose, we rewrite the equation in the following suggestive form,
\bea \label{D}
\mathcal{D} \Delta h_{ij} = 0 ~;\ \ \text{where}\ \mathcal{D} = 1 + \beta \Delta ~.
\eea
Using the short-hand notation $\bar{h}_{ij} := \partial_v h_{ij}$, we can perform a Fourier transform in the $v$-direction,
\bea
\widetilde{h}_{ij}(u, p_v) = \int dv\ e^{-i p_v v}\ \bar{h}_{ij}(u,v) ~,
\eea
which brings the equation of motion in momentum space to the form
\bea \label{FT}
\left[ 1 - 4 i \beta p_v \left( \partial_u + i h_0 p_v \right) \right] \widetilde{K}_{ij}(u, p_v) = 0 ~,
\eea
where $\widetilde{K}_{ij}(u, p_v) = \left( \partial_u + i h_0 p_v \right) \widetilde{h}_{ij}(u, p_v)$. This differential equation in $u$ can be solved to obtain
\begin{align} \label{K}
\widetilde{K}_{ij}(u, p_v, x_i) = \widetilde{K}_{ij}^{(0)}(p_v, x_i)\ \operatorname{exp}\left({\frac{u}{4 i \beta p_v}} \right)
\operatorname{exp} \left({-i p_v \int^{u} d u\ h_{0}\left(u, x_{i}\right)} \right) ~,
\end{align}
where we explicitly inserted back the $x_i$-dependence. Using the definition of $\widetilde{K}_{ij}(u, p_v, x_{i})$, we can now express $\widetilde{h}_{ij}(u, p_v, x_{i})$ in the form given by eq.(\ref{ht}). Finally, we achieve our goal by integrating the equation $\partial_v h_{ij} = \bar{h}_{ij}$ and write down the full solution as given in (\ref{sol}).



\begin{thebibliography}{99}

\bibitem{Adams:2006sv}
A.~Adams, N.~Arkani-Hamed, S.~Dubovsky, A.~Nicolis and R.~Rattazzi,
``Causality, analyticity and an IR obstruction to UV completion,''
JHEP \textbf{10}, 014 (2006)
[arXiv:hep-th/0602178 [hep-th]].

\bibitem{Camanho}
X.~O.~Camanho, J.~D.~Edelstein, J.~Maldacena and A.~Zhiboedov,
``Causality constraints on corrections to the graviton three-point coupling,''
JHEP \textbf{02}, 020 (2016)
[arXiv:1407.5597 [hep-th]].

\bibitem{Hollowood:2016ryc}
T.~J.~Hollowood and G.~M.~Shore,
``Causality, renormalizability and ultra-high energy gravitational scattering,''
J. Phys. A \textbf{49}, 215401 (2016)
[arXiv:1601.06989 [hep-th]].

\bibitem{Chowdhury:2019kaq}
S.~D.~Chowdhury, A.~Gadde, T.~Gopalka, I.~Halder, L.~Janagal and S.~Minwalla,
``Classifying and constraining local four photon and four graviton S-matrices,''
JHEP \textbf{02}, 114 (2020)
[arXiv:1910.14392 [hep-th]].

\bibitem{Chandorkar:2021viw}
D.~Chandorkar, S.~D.~Chowdhury, S.~Kundu and S.~Minwalla,
``Bounds on Regge growth of flat space scattering from bounds on chaos,''
JHEP \textbf{05}, 143 (2021)
[arXiv:2102.03122 [hep-th]].

\bibitem{Edelstein:2016nml} 
J.~D.~Edelstein, G.~Giribet, C.~Gomez, E.~Kilicarslan, M.~Leoni and B.~Tekin,
``Causality in 3D massive gravity theories,''
Phys.\ Rev.\ D {\bf 95}, 104016 (2017)
[arXiv:1602.03376 [hep-th]].\\

\bibitem{Papallo:2015rna}
G.~Papallo and H.~S.~Reall,
``Graviton time delay and a speed limit for small black holes in Einstein-Gauss-Bonnet theory,''
JHEP \textbf{11}, 109 (2015)
[arXiv:1508.05303 [gr-qc]].

\bibitem{Hollowood:2015elj}
T.~J.~Hollowood and G.~M.~Shore,
``Causality violation, gravitational shock waves and UV completion,''
JHEP \textbf{03}, 129 (2016)
[arXiv:1512.04952 [hep-th]].

\bibitem{deRham:2020zyh}
C.~de Rham and A.~J.~Tolley,
``Causality in curved spacetimes: The speed of light and gravity,''
Phys. Rev. D \textbf{102}, 084048 (2020)
[arXiv:2007.01847 [hep-th]].


\bibitem{Mozaffar:2016hmg}
M.~R.~Mohammadi Mozaffar, A.~Mollabashi, M.~M.~Sheikh-Jabbari and M.~H.~Vahidinia,
``Holographic entanglement entropy, field redefinition invariance and higher derivative gravity theories,''
Phys. Rev. D \textbf{94}, 046002 (2016)
[arXiv:1603.05713 [hep-th]].

\bibitem{Deser:2002jk}
S.~Deser and B.~Tekin,
``Energy in generic higher curvature gravity theories,''
Phys. Rev. D \textbf{67}, 084009 (2003)
[arXiv:hep-th/0212292 [hep-th]].\\

\bibitem{Stelle:1976gc}
K.~S.~Stelle,
``Renormalization of higher derivative quantum gravity,''
Phys. Rev. D \textbf{16}, 953 (1977).

\bibitem{Buchbinder:1992rb}
I.~L.~Buchbinder, S.~D.~Odintsov and I.~L.~Shapiro,
``Effective action in quantum gravity", 
IOP, Bristol and Philadelphia (1992).

\bibitem{Nojiri:2001ae}
S.~Nojiri, S.~D.~Odintsov and S.~Ogushi,
``Cosmological and black hole brane world universes in higher derivative gravity,''
Phys. Rev. D \textbf{65}, 023521 (2002)
[arXiv:hep-th/0108172 [hep-th]].

\bibitem{Alvarez-Gaume:2015rwa}
L.~Alvarez-Gaume, A.~Kehagias, C.~Kounnas, D.~L\"ust and A.~Riotto,
``Aspects of quadratic gravity,''
Fortsch. Phys. \textbf{64}, 176 (2016)
[arXiv:1505.07657 [hep-th]].

\bibitem{Salvio:2018crh}
A.~Salvio,
``Quadratic gravity,''
Front. in Phys. \textbf{6}, 77 (2018)
[arXiv:1804.09944 [hep-th]].

\bibitem{Berej:2006cc}
W.~Berej, J.~Matyjasek, D.~Tryniecki and M.~Woronowicz,
``Regular black holes in quadratic gravity,''
Gen. Rel. Grav. \textbf{38}, 885 (2006)
[arXiv:hep-th/0606185 [hep-th]].

\bibitem{Matyjasek:2004vr}
J.~Matyjasek and D.~Tryniecki,
``Charged black holes in quadratic gravity,''
Phys. Rev. D \textbf{69}, 124016 (2004)
[arXiv:gr-qc/0402098 [gr-qc]].

\bibitem{Svarc:2018coe}
R.~Svarc, J.~Podolsky, V.~Pravda and A.~Pravdova,
``Exact black holes in quadratic gravity with any cosmological constant,''
Phys. Rev. Lett. \textbf{121}, 231104 (2018)
[arXiv:1806.09516 [gr-qc]].

\bibitem{Audretsch:1993kp}
J.~Audretsch, A.~Economou and C.~O.~Lousto,
``Topological defects in gravitational theories with nonlinear Lagrangians,''
Phys. Rev. D \textbf{47}, 3303 (1993)
[arXiv:gr-qc/9301024 [gr-qc]].

\bibitem{Cao:2013osa}
Z.~Cao, P.~Galaviz and L.~F.~Li,
``Binary black hole mergers in $f(R)$ theory,''
Phys. Rev. D \textbf{87}, 104029 (2013)
[arXiv:1608.07816 [gr-qc]].

\bibitem{Ghosh:2019twk}
A.~Ghosh, S.~Jana, A.~K.~Mishra and S.~Sarkar,
``Constraints on higher curvature gravity from time delay between GW170817 and GRB 170817A,''
Phys. Rev. D \textbf{100}, 084054 (2019)
[arXiv:1906.08014 [gr-qc]].

\bibitem{'t Hooft}
T.~Dray and G.~'t Hooft,
``The gravitational shock wave of a massless particle,''
Nucl. Phys. B \textbf{253}, 173 (1985).

\bibitem{Sexl}
P.~C.~Aichelburg and R.~U.~Sexl,
``On the gravitational field of a massless particle,''
Gen. Rel. Grav. \textbf{2}, 303 (1971).

\bibitem{Campanelli}
M.~Campanelli and C.~O.~Lousto,
``Exact gravitational shock wave solution of higher order theories,''
Phys. Rev. D \textbf{54}, 3854 (1996)
[arXiv:gr-qc/9512050 [gr-qc]].

\bibitem{Abe:2017abx}
Y.~Abe, T.~Inami, K.~Izumi and T.~Kitamura,
``Matter scattering in quadratic gravity and unitarity,''
PTEP \textbf{2018}, no.3, 031E01 (2018)
[arXiv:1712.06305 [hep-th]].

\bibitem{Brigante:2007nu}
M.~Brigante, H.~Liu, R.~C.~Myers, S.~Shenker and S.~Yaida,
``Viscosity bound violation in higher derivative gravity,''
Phys. Rev. D \textbf{77}, 126006 (2008)
[arXiv:0712.0805 [hep-th]].

\bibitem{CEZ}
X.~O.~Camanho, J.~D.~Edelstein and A.~Zhiboedov,
``Weakly coupled gravity beyond general relativity,''
Int. J. Mod. Phys. D \textbf{24}, 1544031 (2015).

\bibitem{Horowitz}
G.~T.~Horowitz and N.~Itzhaki,
``Black holes, shock waves, and causality in the AdS/CFT correspondence,''
JHEP \textbf{02}, 010 (1999)
[arXiv:hep-th/9901012 [hep-th]].

\bibitem{Benakli:2015qlh}
K.~Benakli, S.~Chapman, L.~Darm\'e and Y.~Oz,
``Superluminal graviton propagation,''
Phys. Rev. D \textbf{94}, 084026 (2016)
[arXiv:1512.07245 [hep-th]].

\bibitem{Kabat}
D.~N.~Kabat and M.~Ortiz,
``Eikonal quantum gravity and Planckian scattering,''
Nucl. Phys. B \textbf{388}, 570 (1992)
[arXiv:hep-th/9203082 [hep-th]].

\bibitem{Giddings:2011xs}
S.~B.~Giddings,
``The gravitational S-matrix: Erice lectures,''
Subnucl. Ser. \textbf{48}, 93 (2013)
[arXiv:1105.2036 [hep-th]].

\bibitem{Maldacena:2015waa}
J.~Maldacena, S.~H.~Shenker and D.~Stanford,
``A bound on chaos,''
JHEP \textbf{08}, 106 (2016)
[arXiv:1503.01409 [hep-th]].

\bibitem{Tekin:2016vli}
B.~Tekin,
``Particle content of quadratic and $f(R_{\mu\nu\sigma \rho})$ theories in $(A)dS$,''
Phys. Rev. D \textbf{93}, 101502 (2016)
[arXiv:1604.00891 [hep-th]].

\bibitem{Caron-Huot:2021enk}
S.~Caron-Huot, D.~Mazac, L.~Rastelli and D.~Simmons-Duffin,
``AdS bulk locality from sharp CFT bounds,''
[arXiv:2106.10274 [hep-th]].

\end{thebibliography}
\end{document}